\documentclass[aps,pra,twocolumn,byrevtex,showpacs,floatfix,superscriptaddress,nofootinbib]{revtex4-1}
\pdfoutput=1
\usepackage{graphicx}
\usepackage{dcolumn}
\usepackage{amsmath}
\usepackage{epsfig}
\usepackage{bm}
\usepackage{amssymb}
\usepackage[normalem]{ulem}
\usepackage{hyperref}
\hypersetup{colorlinks=true,linkcolor=blue,citecolor=blue,urlcolor=blue} 
\usepackage[T1]{fontenc}
\usepackage{courier}
\usepackage{cleveref}
\usepackage{soul}
\usepackage{cancel}

  \newcommand{\ket}[1]{|#1\rangle}
  \newcommand{\bra}[1]{\langle #1|}
  \newcommand{\braket}[2]{\langle #1\mid #2\rangle}

  \DeclareMathOperator{\Tr}{Tr}

\begin{document}
  
\title{Classical Limits and Contextuality in a Scenario of Multiple Observers}

\author{Roberto D. Baldij\~ao}
\affiliation{Institute of Physics ``Gleb Wataghin'', State
University of Campinas 
13083-859, Campinas, SP, Brazil }
\affiliation{Institute for Quantum Optics and Quantum Information, Austrian Academy of Sciences, Boltzmanngasse 3, 1090 Vienna, Austria}

\author{Marcelo Terra Cunha}
\affiliation{Instituto de Matem\'atica, Estat\'istica e Computa\c{c}\~{a}o Cient\'ifica, Universidade Estadual de Campinas, Campinas, SP, Brazil.}

\date{\today}

\begin{abstract}
Contextuality is regarded as a non-classical feature, challenging our everyday intuition; quantum contextuality is currently seen as a resource for many applications in quantum computation, being responsible for quantum advantage over classical analogs. In our work, we adapt the $N$-cycle scenarios with odd $N$ to multiple independent observers which measure the system sequentially. We analyze the possibility of violating the inequalities as a function of the number of observers and under different measurement protocols. We then reinterpret the results as an open quantum system where the environment is divided into fragments. In this context, the results show the emergence of non-contextuality in such a setting, bringing together the quantum behavior to our classical experience. We then compare such emergence of non-contextuality with that of {objectivity under the `Environment as a Witness paradigm'.}

\end{abstract}

\maketitle

\section{Introduction}
\label{sec: Intro}

Quantum Contextuality, i.e.~the failure of explaining the probabilities obtained by Quantum Theory (QT) as a result of unknown well defined answers for every question allowed by a specified scenario, is a remarkable non-classical treat. Indeed, in Classical Theories (CTs), non-contextuality is a natural assumption \cite{Spekkens1,SpekkensTalk} and these theories are very successful in explaining our everyday experience. 
This incompatibility was first shown as a logical impossibility between any theory with such a feature and the predictions of QT \cite{KS}. 
{The first and stronger form of quantum contextuality is known as state-independent contextuality, when a collection of measurements satisfying some constraints generates correlations which cannot be explained under non-cintextuality assumption for any quantum state \cite{KS,Peres1990_Square,Mermin1990_Square}.
A weaker but very interesting form of contextuality happens when the contradiction appears for some quantum states \cite{Bell64,KCBS}. This is called state-dependent (quantum) contextuality and it is the form of contextuality we use in this Article.}
A very practical way of witnessing contextuality is using non-contextuality inequalities \cite{KCBS,JA02}, { which are always obeyed by any non-contextual theory, but possibly violated by contextual theories, such as QT.
Suitable choices of quantum states and measurements lead to violations of these inequalities, as verified in the laboratory \cite{MauricioeCanas,Ahrens2013,KirchmairBlattC,LeupoldPRL,GIL}.
In the lab, one experimentalist chooses randomly among the possible contexts of the scenario, or among the terms involved in the inequality, and makes the measurements, recording the data for future processing.}

One interesting open question concerns the robustness of quantum contextuality to sequential \emph{independently chosen} measurements on the same system, made by different observers. 
Analogous questions were discussed for scenarios regarding steering \cite{Sasmal2018} and non-locality \cite{Silva2015,Das2019}, but a similar study on contextuality-like setups is still lacking. 
It is necessary to understand the limits that quantum correlations obey in a setting that does not necessarily respect non-disturbance (or non-signaling in a non-locality setting) \cite{Das2019,Silva2015}, since the sequential measurements are not necessarily compatible.  
This has both a foundational importance and  a practical one. In practical terms this non-classical feature is a resource for quantum to overcome classical computation \cite{Karanjai2018,HowardCompQC}, {and} analyzing how robust it is for sequential measurements might be interesting for some algorithms that use an analogous sequential setup \cite{HowardCompQC,MBQC-Contextuality,MBQC,MBQC-Review}.
Foundationally, it touches the question of how we, observers, create some classical picture of what is usually called \emph{reality} via a process within QT; this classical limit being attained here by the absence of violation of non-contextuality inequalities, which is consistent with a probabilistic view of the world based on the lack of precise information of some idealized underlying reality.

In this work, we discuss the robustness of contextuality for independent observers using the odd $N$-cycle scenarios \cite{N-ciclo}.
Those scenarios are interesting since they are the building blocks for contextuality \cite{Cabello2010,CSW}. 
Moreover, they cannot be mapped into any bipartite scenario, revealing an essential picture of contextuality. 
{We introduce and study three different measurement protocols {which,}
despite being  equivalent to test contextuality in the usual single observer implementation,  behave quite differently in this multiplayer setting.} {Analyzing for the best realization for the single observer case,} we obtain the best quantum results for each player in each protocol.
{These results show} that contextuality quickly disappears in most cases. 
For one specific protocol, however,  violations of non-contextuality inequalities can be found by more than one observer, which can be viewed as a way of protecting quantum contextuality.
In this sense, the generic result is consistent with the emergence of a classical world from the quantum realm, while the peculiar exception is a guide towards the quantum engineering and protection of contextual machines.

{We move to interpret these results in the light of an interaction of the system and a fragmented environment, called by Zurek\textit{ et al.} as the paradigm of `Environment as a Witness' or the `Redundancy program' \cite{ZUREKREVIEW2007,Ollivier_QD}. {In such an approach, the environment is understood as a channel broadcasting information about the system. 
Following this paradigm, some processes to explain how objectivity might emerge were proposed: {Quantum Darwinism (QD) \cite{ZUREKQDEnvariance,ZUREK2003,Ollivier_QD}, State Spectrum Broadcasting (SSB)\cite{RHorodecki_SSB} and, finally, Strong Quantum Darwinism (SQD) -- the latter providing sufficient and necessary conditions for emergence of objectivity\cite{Le2019_SQD}.} These processes differ in important aspects \cite{RHorodecki_SSB,Le2019_SQD}, but they do agree on the concept of objectivity: it is attained when independent observers obtain the same information about the system. 
One very beautiful consequence of this concept is that, when attained, the information about that specific observable of the system is redundantly available in the environment, allowing anyone to learn about it, without disturbing the system.
Another very appropriate name for this kind of `objectivity' through agreement of independent agents is \emph{inter-subjectivity} \cite{Mironowicz2017}.
From this perspective,  our results say that, for all practical purposes, odd $N$-cycle \emph{non-contextuality} emerges from the quantum realm, even in this well designed environment.}}
This indicates that non-contextuality is a classical feature that can emerge in  open quantum systems.

This paper is organized as follows: in Sec. \ref{Sec: ScenariosandMore} we review some important features of the $N$-cycle scenarios and introduce three possible protocols a single player can use to test for contextuality in these scenarios. In Sec. \ref{Sec: Approach} we develop the multiplayers scenario and expose the results. In Sec. \ref{Sec: QD} we interpret these results as emergence of non-contextuality under system-environment interactions, discussing its similarities and differences with emergence of {objectivity under the `Environment as a Witness' paradigm}. Conclusion and open questions are discussed in Sec. \ref{Sec: Conclusions}.
The actual proofs of the results are detailed in appendices.

\section{Scenarios, Inequalities and Protocols} 
\label{Sec: ScenariosandMore}

Kochen-Specker contextuality scenarios are constituted by a set of available observables (or measurements), the compatibility restrictions among them and the set of outcomes \cite{CSW,TerraBarbaraBook}. 
The compatibility restrictions can be depicted in a compatibility graph, where each vertex represents an observable and two vertices are connected if and only if they represent compatible observables; sets where all vertices are pairwise connected are called contexts. 
In the $N$-cycle scenarios, we have $N$ observables and the compatibility graph is a cycle of length $N$  \cite{N-ciclo}. 
In other words, denoting $A_i$ the observables, the maximal contexts are the elements in the set $\{\{A_0,A_{1}\},...,\{A_{N-1},A_{0}\}\}$. The observables are dichotomic with outcome events $o_i\in\{-1,1\}$ and $i\in\{0,1,...,N-1\}$ labelling each measurement.
It is usual to consider the correlations given by theories that respect the \emph{non-disturbance principle}, which states that marginalizing joint probabilities of outcomes of compatible measurements always reproduce the probabilities for the restricted context: $\sum_{o_j}p(o_i,o_j|i,j) = p(o_i|i)$ for every measurement $j$ compatible with $i$.

The non-trivial inequalities defining the classical polytope for odd $N$, after relabelling outcomes if necessary, can be written as
\begin{equation}
    \sum_{i=0}^{N-1} \langle A_iA_{i+1}\rangle \, \stackrel{\rm NC}{\geq}\, 2-N,
    \label{Ineq: Correlators}
\end{equation}
with sums in the indexes made modulo $N$ \cite{N-ciclo}.
Considering our choice of outcomes, the avarage value is given by
{
 \begin{equation}
 \langle A_iA_{i+1}\rangle = p(o_i=o_{i+1}|i,{i+1}) - p(o_i\neq o_{i+1}|i, {i+1}). 
 \end{equation}}
 The superscript ``NC'' in Ineq, ~\eqref{Ineq: Correlators} reminds that the inequality was derived assuming non-contextuality. This means that the deterministic outcome assigned to a measurement does not depend on which context it was measured, e.g., if one measures $\{A_0,A_1\}$ or $\{A_1,A_2\}$, a non-contextual theory must assign the same $o_1$ irrespective of the context and so on.
Inequality \eqref{Ineq: Correlators} just means that if we try to assign dichotomic values to all observables non-contextually, at least two neighbors must agree.
Another way of saying this is through inequality
\begin{subequations}
\begin{equation}
    \sum_{i=0}^{N-1}p(o_i = o_{i+1}|i,i+1)  \stackrel{\rm NC}{\geq} 1, 
    \label{Ineq: ProbEqual}
\end{equation}
which, through the normalization condition $p(o_i=o_{i+1}|i,i+1) + p(o_i\neq o_{i+1}|i,i+1) = 1$, is equivalent to
\begin{equation}
      \sum_{i=0}^{N-1}p(o_i\neq o_{i+1}|i,i+1) \stackrel{\rm NC}{\leq}  N-1.
    \label{Ineq: ProbDif}
\end{equation}
\label{Ineq: Prob}
\end{subequations}

Now, if we assume \emph{exclusiveness} \cite{Cabello2010}, that is, 
assume that $p(-1,-1|i,i+1)=0$ for all contexts, then the joint measurements become three-outcome measurements: the event $(-1,-1|i,i\pm1)$ never happens.
Then, $\sum_{o}p(-1,o|i,j)$ reduces to $p(-1,+1|i,j)$ and non-disturbance reads $p(-1,+1|i,j) \equiv p(-1|i)$ for all measurements $A_j$ compatible with $A_i$. 
So, in a theory obeying non-disturbance $+$ exclusiveness in this dichotomic scenario, we can identify, {counterfactually}, the events $(-1,+1|i,i+1) \leftrightarrow (+1,-1|i-1,i) \leftrightarrow (-1|i)$.
This allows us to write:
\begin{equation}
     p(o_i\neq o_{i+1}|i,i+1)  = p(-1|i)+p(-1|i+1). 
    \label{Eq: NDplusExcl}
\end{equation}

\begin{figure}[ht]
    \includegraphics[scale=0.28]{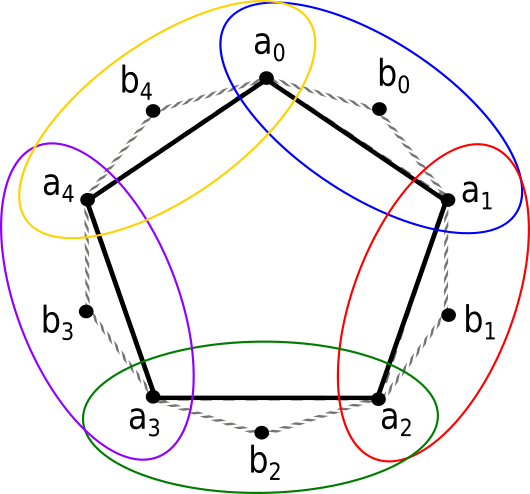}
    \caption{Hypergraph representing a $5$-cycle realization in a theory obeying $p(-1,-1|i,i+1)=0$. Hyperedges represent each context $\{A_i,A_{i+1}\}$. The $a_i$ vertices denote the outcomes `yes' to the related effects, important for expression $\alpha$ (hence the continuous line) and the $b_i$ vertices are `yes' answers to the related effects, where the dashed line reinforce its exclusiveness only to $\{a_i,a_{i+1}\}$. Non-disturbance in this dichotomic scenario, or the validity of the `Gleason property' for the effects, allows us to consider $a_i$ `the same' vertex for both hyperedges it belongs to (see \ref{Sec: NDandGP}). Non-contextuality to deterministic assignments then implies Inequalities \eqref{Ineq: ProbAeB}.}
    \label{Fig: HyperGraph}
\end{figure}

Let us now assume a generalized probabilistic theory framework to describe how probabilities are obtained \cite{Barret2007,Janotta2014}. 
In a scenario with non-disturbance, exclusiveness, and dichotomic measurents, 
one can associate to the event `measuring $A_i$ and obtaining output $o_i=-1$ (resp.~ $o_i=1$)' the effect $a_i$ ($\neg {a_i}$) and to the  joint measurement of $\{A_i,A_{i+1}\}$  the set of three exclusive effects $\{a_i \wedge \neg a_{i+1}, \neg a_i \wedge a_{i+1}, \neg a_i \wedge \neg a_{i+1}\}$.
The probabilities of the respective events are given by the expected values of such effects\footnote{The identification of events that led to Eq. \eqref{Eq: NDplusExcl} is translated to $a_i\wedge \neg a_{i+1} \leftrightarrow a_i\wedge \neg a_{i-1} \leftrightarrow a_i$. This last identification between effects is also related to the stronger principle of `Gleason Property'\cite{Cabello2010,Gleason}. 
See the Appendix \ref{Sec: NDandGP} for a quick exposition about the relation and differences between this property, non-disturbance and their equivalent consequences in this scenario.}.  
In terms of effects, we can rewrite Eq. \eqref{Eq: NDplusExcl}:
\begin{align}
    &p(o_i\neq o_{i+1}|i,i+1) \nonumber
    \\
    &= p(-1|i)+p(-1|i+1) =\left\langle a_i \right\rangle + \left\langle a_{i+1} \right\rangle.
    \label{Eq: NDplusExclEffects}
\end{align}
The probabilities of disagreement can then be written only in terms of $\left\{\langle a_i \rangle\right\}$.
The agreement probabilities, because of exclusiveness, depends only on the expected values of $\{\neg a_i \wedge \neg a_{i+1}\}$. 
For a lighter notation, and for historical reasons \cite{Wright}
, we will denote $b_i$ instead of $\neg a_i \wedge \neg a_{i+1}$ (see also Fig. \ref{Fig: HyperGraph}). 
Then, Ineqs. \eqref{Ineq: Prob} turn (with a small twist in ordering) into: 
\begin{subequations}
\begin{align}
    \alpha = \sum_{i=0}^{N-1}\left\langle a_i\right\rangle \,&\stackrel{\rm NC}{\leq} \,\, \frac{N-1}{2},
    \label{Ineq: ProbA}\\
    \beta = \sum_{i=0}^{N-1}\left\langle b_i \right\rangle \, &\stackrel{\rm NC}{\geq}\,\, 1. 
    \label{Ineq: ProbB}
\end{align}
\label{Ineq: ProbAeB}
\end{subequations}

These are the Inequalities that will be used from now on (Inequality \eqref{Ineq: ProbA} for $N=5$  is the KCBS inequality \cite{KCBS}).  
It is important noticing that violation of one inequality of \eqref{Ineq: ProbAeB} leads to violation of the other, and so they are equivalent with respect to witnessing contextuality. 
Another important comment is that, surprisingly, Inequality \eqref{Ineq: ProbA} shows explicitly that we can look to the results of a measurement of each $A_i$ alone, instead of a whole context $\{A_i,A_{i+1}\}$, under the validity of the exposed assumptions. 
Of course, experimentally, it is important to show that such conditions are met \cite{Cabello2010}. 
Fig. \ref{Fig: HyperGraph} exhibit the pentagon scenario under the exclusiviness assumption, which gives rise to the trichotomic measurements with effects $\left( a_i, b_i, a_{i+1} \right)$.

The trichotomic measurements  $\left( a_i, b_i, a_{i+1} \right)$ together with Inequalities \eqref{Ineq: ProbAeB} suggest three different protocols for testing for contextuality in the $N$-cycle obeying exclusiveness hypothesis.
For each value of $i$, the first protocol specify the complete measurement of this hyperedge, while two other protocols come from two natural {coarse-grainings}/dichotomizations, i.e:
\begin{enumerate}
    \item $\mathcal{M}_i = \{a_i,b_i,a_{i+1}\}$; suitable for both Inequalities;
    \item $\mathcal{M}^a_i=\{a_i,\neg a_i\}$ suggested by Ineq. \eqref{Ineq: ProbA};
   \item  $\mathcal{M}^b_i=\{b_i,\neg b_i\}$, suggested by Ineq. \eqref{Ineq: ProbB}.
\end{enumerate}
Protocol $1$ is motivated by the usual contextuality scenario, where joint or sequential measurements of the context allows to recover
separately the results for both observables in that context. 
However, as  Inequalities \eqref{Ineq: ProbAeB} show, this can carry excess of information: under the assumption $p(-1-1|i,i+1)=0$, one (surprisingly) might know only the occurrences of $a_i$ or $b_i$, which are less invasive measurements (used in several experimental tests of similar inequalities \cite{KirchmairBlattC,MauricioeCanas,Ahrens2013}). 
In a typical setup with a single observer this makes no difference (since after measuring one context the system is prepared again for another round). 
This is not the case for sequential observers, as we shall explain soon.
Protocol $2$ is designed to evaluate the quantity $\alpha$ and test contextuality through Ineq.~\eqref{Ineq: ProbA}, while protocol $3$ focus on $\beta$ and Inequality \eqref{Ineq: ProbB}.

Since our motivation is to consider the robustness of quantum contextuality, it is important to look closely to quantum realizations of the odd $N$ scenarios.
The maximum quantum violations for all odd $N\geq 5$ can be realized in a Hilbert space of dimension $d=3$, with measurements defined by $A_i = \mathcal{I} - 2\ket{a_i}\bra{a_i}$, where $\mathcal{I}$ is the identity and, with an appropriate basis choice, the vectors $\ket{a_i}$ are
\small
    \begin{equation}
     \ket{a_i} =\mathcal{K}\left(\cos{\left(\frac{i\pi(N-1)}{N}\right)},\sin{\left(\frac{i\pi(N-1)}{N}\right)},\sqrt{\cos{\left(\frac{\pi}{N}\right)}}\right)^t \nonumber,
    \label{Eq: Projectors}
    \end{equation}
    \normalsize
with $t$ meaning transposition, and $\mathcal{K} =1/\sqrt{(1+\cos(\pi/N))}$ for normalization \cite{N-ciclo}. The vectors $\ket{b_i}$ are orthogonal to $\{\ket{a_i},\ket{a_{i+1}}\}$. 
The important point here is the symmetry obeyed by the vectors $\{\ket{a_i}\}$: they form a regular polygon in a plane orthogonal to the axis $(0,0,1)^t$. The state that reaches the maximum violations is $\ket{\psi_{handle}} = (0,0,1)^t$, symmetric with respect to the $\{\ket{a_i}\}$. 
The form above for the vectors implies $\braket{a_i}{a_{i\pm1}} = 0$, obeying the compatibility and exclusiveness constraints required by the scenario. 

\section{Multiplayers Sequential Approach}
\label{Sec: Approach}

Usual tests of contextuality involve only one ``player''.
Even if many experimentalists access the same system, they do it as a coordinate team.
The multiplayer sequential approach (MPSA)  that we study here goes in the opposite way: the same quantum system will be tested by different players.
The players agree on which protocol to use, but they independently choose the measurement to implement without any knowledge related to other players' measurements or results.

{By comparison, if an experiment is draw to test inequality \eqref{Ineq: ProbA} in a given $N$-cycle scenario, usually one considers identical preparations of a system; at each run, the experimentalist chooses $i$, applies the measurement $\mathcal{M}^a_i$, collects the result and discards the system.
In the MPSA, given a prepared system, the first player chooses $i_1$, applies the measurement $\mathcal{M}^a_{i_1}$, collects the result and leaves the system for the next player, who independently chooses $i_2$ and so on.}

{The question we focus is whether the $k$-th player will be able to witness contextuality under this rules.
We will see that the answer depends strongly on the choice of the protocol to be used and of the inequality to be tested.}

\begin{figure}[ht]
    \includegraphics[scale=0.36]{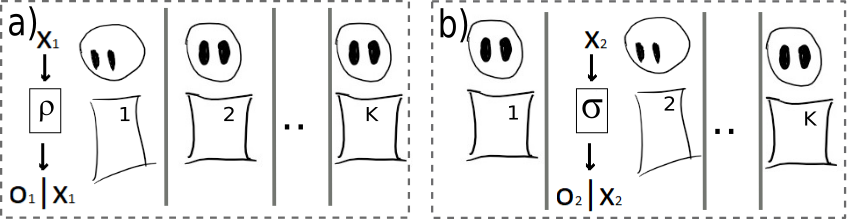}
    \caption{Scheme for the multiplayers game. a) First player makes a measurement $i_1$ on state $\rho$, obtaining his outcome $o_{i_1}|i_1$. All other players ignores $i_1$ and $o_{i_1}$. b) Second player makes a measurement on state $\sigma$ which depends on $\rho$, $i_1$ and $o_{i_1}|i_1$, and so on.}
    \label{Fig: Game}
\end{figure}

{More precisely,} the game to analyze the resistance of quantum contextuality under sequential independent players is formulated as follows (see Fig. \ref{Fig: Game}). 
Fix a $N$-cycle scenario, an Inequality of \eqref{Ineq: ProbAeB} (which all players will evaluate), and a related protocol. Then, (i) an initial state of dimension $d=3$ is prepared; (ii) an order of access of each player to $\mathcal{S}$ is followed;   (iii) each player chooses one of the possible quantum projective measurements and executes it on $\mathcal{S}$.  
Steps (i) to (iii) (called henceforth a run) are repeated for players to collect their individual data to estimate the expression $\alpha_Q^k$ or $\beta_Q^k$, where superscript $k$ labels each player and $Q$ reminds that we are looking at quantum realizations.
Player $k$ wins the game if capable of violating the chosen inequality. 
Players cannot communicate, being ignorant with respect to others' measurements and outcomes\footnote{{This is the sense of saying we relax the non-disturbance condition: the measurements and inequality evaluation of each player dobey non--disturbance, but what was done with the state by previous players will generally not be compatible to what will be done in the future.}}. The game is represented in figure \ref{Fig: Game}; in the following, we analyze the behavior of the observables involved in the inequalities in each of the protocols. 

\begin{figure}[h]
    \includegraphics[scale=0.24]{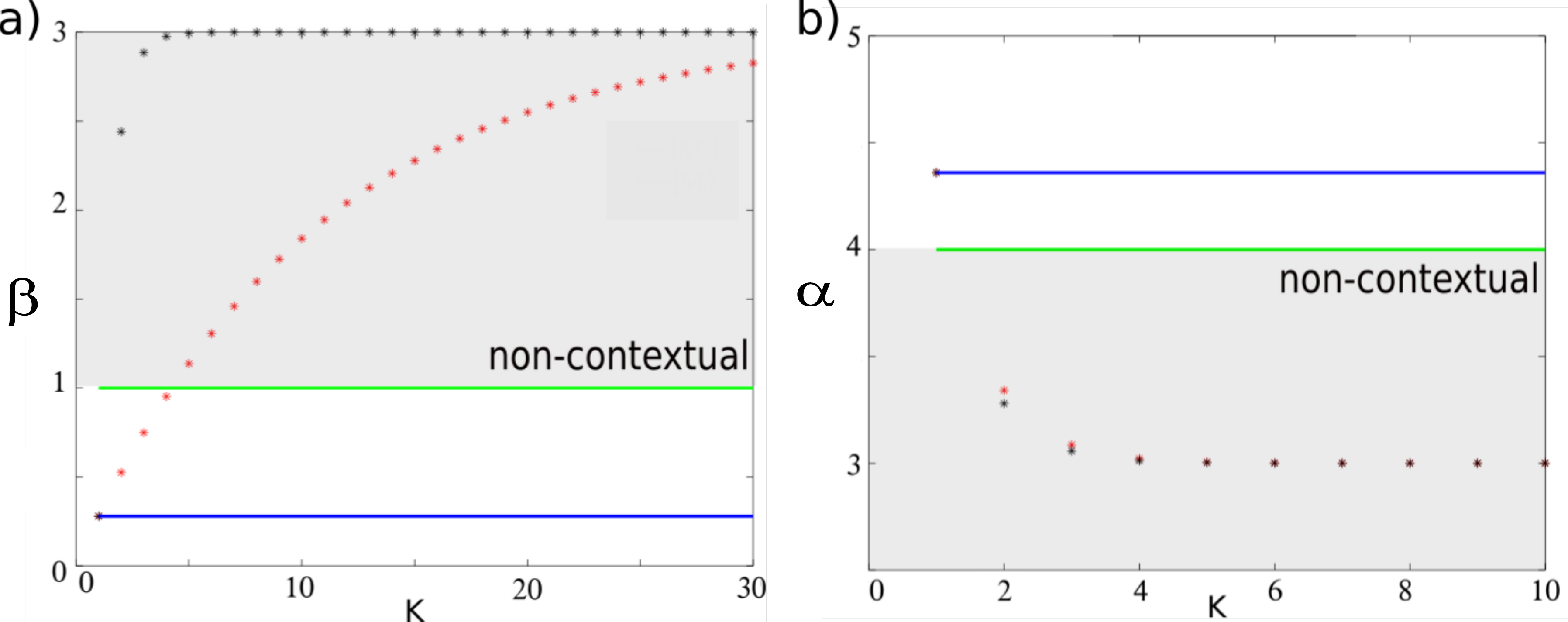}
    \caption{a) minimum value of $\beta_Q^k$ for the $k$-th player, for protocols $1$ (black) and $3$ (red). b) maximum value of $\alpha_Q^k$ for the $k$-th player for measurement protocols $1$ (black) and $2$ (red). Initial state is $\ket{\psi_{handle}}$ and $N=9$. Blue line sets the quantum highest violation and Green line sets the non-contextual bounds. 
    }
    \label{Fig: ViolationProtocols}
\end{figure}

For protocol $1$, it is useful to define the probability vector $\vec{P}_{i_k}$: each entry tells the probability of obtaining each outcome for the measurement $\mathcal{M}_{i_k}$, with $i_k\in\{0,...,N-1\}$ labelling the choice of the $k$-th observer. For instance, suppose $i_k=3$, then $P_{i_k=3}=(p(a_3|3),p(b_3|3),p(a_4|3))^t$. With $(\vec{P}_{i_k})_{i_k}$ at hand, each observer can calculate either $\alpha_Q$ or $\beta_Q$ by choosing the relevant entries, i.e., using the relation
\begin{subequations}
\begin{equation}
    \alpha_Q^k = \sum_{i_k}\langle v_{\alpha},\vec{P}_{i_k}\rangle = \langle v_{\alpha},\sum_{i_k}\vec{P}_{i_k}\rangle,
    \label{Eq: alphaProtocolOne}
    \end{equation}
    \begin{equation}
        \beta_Q^k = \sum_{i_k}\langle v_{\beta},\vec{P}_{i_k}\rangle = \langle v_{\beta},\sum_{i_k}\vec{P}_{i_k}\rangle,
        \label{Eq: betaProtocolOne}
\end{equation}
\label{Eq: IProtocolOne}
\end{subequations}
where $\langle \cdot,\cdot \rangle$ is the usual Euclidean scalar product, $v_{\beta}^{t} = (0,1,0)$ and $v_{\alpha}^{t} = (1/2,0,1/2)$, see \ref{Sec: Protocol1}.
Since protocol $1$ is composed of complete measurements, each of its elements prepares completely the state after a measurement. Hence the dynamics under this protocol takes the form of a Markovian process on the post-measured states (details in Sec. \ref{Sec: Protocol1}). Considering that each player chooses uniformly which measurement to perform and that the same order of players is used in every run, it is possible to show that
\begin{align}
    \sum_{i_k} \vec{P}_{i_k} = \sum_{i_{k-1}}\mathbb{M}_N\vec{P}_{i_{k-1}} 
    = \left(\mathbb{M}_N\right)^{k-1} \sum_{i_1}\vec{P}_{i_1},
    \label{Eq: Markov}
\end{align}
where $\mathbb{M}_N$ is a bistochastic matrix. Given the quantum realization described above, $\mathbb{M}_N$ depends only on the $N$-cycle scenario being tested (thus is fixed). It could happen that the initial state allowing for the best violation for the $k$-th observer would depend on $k$. However Eq. \eqref{Eq: Markov}, together with the form of $\mathbb{M}_N$, shows that the initial quantum state which reaches the maximum violation for the first player maximizes (or minimizes) the value of $\alpha_Q^k$ ($\beta_Q^k$) for the $k$-th observer (see Sec. \ref{Sec: Protocol1} for details). Eq.s \eqref{Eq: Markov} also permits to obtain all the values $(\alpha^k_Q)_k$ or $(\beta_Q^k)_k$ and also to calculate the asymptotic limit.  
Because $\mathbb{M}_N$ represents a regular (and so, irreducible) Markov process for every odd $N$, its steady eigenvector $\vec{P}^*$ is unique \cite{DurrettStochasticBook}. Since $\mathbb{M}_N$ is bistochastic this is given by the uniform distribution 
$\vec{P}^*_i = 1/3$, which implies
\begin{equation}
    \lim_{k \to \infty} \alpha_Q^k =\lim_{k \to \infty} \beta_Q^k= \frac{N}{3}.
    \label{Eq: AsymptoticMarkov}
\end{equation}
Analyzing $\alpha^k_Q$ and $\beta_Q^k$ for each $k$ via Eq. \eqref{Eq: IProtocolOne}, we see that there is no violation already for the second observer, for all $N$ (see Table \ref{Table1} and Fig. \ref{Fig: ViolationProtocols}). This is a rather extreme behavior, leading to impossibility of attesting non-contextuality after measurement on $\mathcal{S}$ by just one player. This can be considered as a consequence of the high level of disturbance of these measurements, since they completely destroy coherences in the measured basis. It is natural to ask if this radical emergence of no-violation also happens for the other protocols, which preserve coherence in a $2$-dimensional subspace of the Hilbert space.

For protocols $2$ and $3$, the post-measurement state can not be completely determined by a previous measurement. Remarkably, using the symmetries of the odd $N$-cycle quantum realizations (defined by the $C_{Nv}$ point group \cite{InuiTanabeOnoderaBook}), it is possible to obtain a relation between $\alpha_Q^k$ and $\alpha_Q^{k-1}$, as well for $\beta^Q$ (see \ref{Sec: Protocols2and3}). Explicitly:
\begin{subequations}
\begin{align}
    \alpha_Q^k &= C_N\alpha_Q^{k-1} +c_N,
    \label{Eq: IneqProtocol2}\\
     \beta_Q^k &= B_N\beta_Q^{k-1}+b_N,
    \label{Eq: IneqProtocol3}
    \end{align}
\label{Eq: IneqProtocols2e3}
\end{subequations}
with $B_N,b_N,C_N,c_N$ fixed for each $N$. With Eqs. \eqref{Eq: IneqProtocols2e3}, it is possible to see again that
the initial state allowing for maximum violations for $k=1$ reaches the highest violations for any $k$, i.e., $\ket{\psi_{\text{handle}}}$. With the form of Eqs. \eqref{Eq: IneqProtocols2e3} it is possible to calculate the quantities for each player and to obtain the asymptotic limit analytically, using the symmetries and dependencies of $B_N\, (C_N)$ and $b_N\,(c_N)$ with $N$ (see \ref{Sec: Protocols2and3}). This gives
\begin{equation}
    \lim_{k \to \infty}\alpha_Q^k= \lim_{k \to \infty} \beta_Q^k=\frac{N}{3},
    \label{Eq: AsymptoticNonMarkov}
\end{equation}
which again means no violation of any of the inequalities and show that dynamics imposed by the game leads to a limit considerably above (below) the non-contextual bound (equal to the asymptotic limit for
protocol $1$). Results presented above
are depicted in Fig. \ref{Fig: ViolationProtocols} for the case $N=9$. Other cases are presented at Table \ref{Table1}.

The striking difference between protocols $2$ and $3$ shows that coherence is not the only ingredient necessary to extend survival of contextuality. This difference must be a consequence of orthogonality in the $\{\ket{a_i}\}$ set: if a player obtains outcome $a_i$, it is automatically forbidden for the next one to obtain $a_{i+1}$ or $a_{i-1}$, while the $\ket{b}$ vectors are less restrictive, since obtaining an outcome $b_i$ implies (with high probability) less disturbance on the incoming state.

We can understand this disappearance of the violations as a consequence of the degradation of the state of the system in each step. Since the inequalities in consideration are state-dependent, this modification of the state makes it less capable of violating the inequalities, until it goes under (above) the classical bound. The average limit of $N/3$ for $K\rightarrow \infty$ is due to the fact that the average state in this limit is in fact the maximally mixed \ref{Sec: AssympState}.

It is noteworthy that independence of players is crucial since \emph{collective strategies can guarantee that all players win}. For instance, if all players combine which measurements to make 
or post-select the data to keep only those not perturbed by precedent players, (maximal) violation is always reachable.

\section{System-Environment Picture and Emergent Classicality}
\label{Sec: QD}

\subsection{Open Quantum System Interpretation}

One valid analysis of those results is obtained by using an open quantum system perspective.
Since the system interacts with the players, they can also be interpreted as an environment, causing the system to decohere, to approach a classical description and to lose contextuality.
However, it is much more informative to interpret such interaction under the so called `Redundancy program' or `Environment as a Witness' paradigm \cite{ZUREKREVIEW2007,Ollivier_QD}. 
Usually in this paradigm, the environment consists of several fragments; independent observers can then learn something about the central system, $\mathcal{S}$, by measuring disjoint sets of such fragments.

\begin{figure}[ht]
    \includegraphics[scale=0.3]{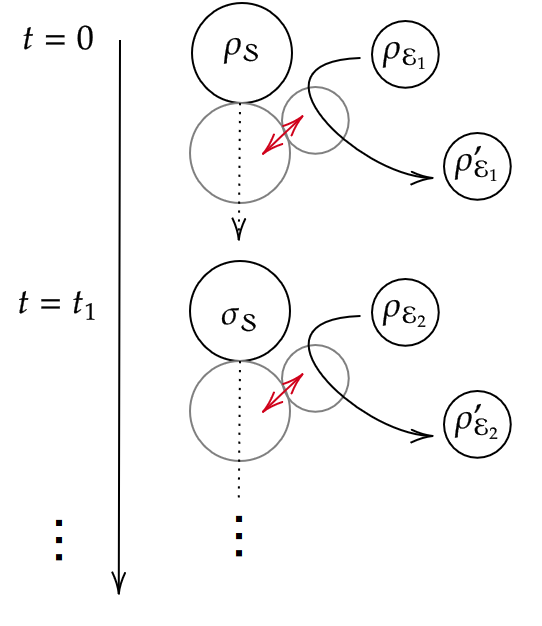}
    \caption{{We can interpret each measurement on the sequential setup as a strong interaction in a collisional approach. In each interaction (e.g. with $\mathcal{E}_1$), an observable is monitored on $\mathcal{S}$ and such information is transfered to the environmental fragment through the correlations established by the interaction; i.e., it can be recovered by measuring $\rho'_{\mathcal{E}_1}$. The decohered $\mathcal{S}$ then interacts with the next system, e.g. $\mathcal{E}_2$ and so on.}}
    \label{Fig: Interaction}
\end{figure}

{Taking this paradigm into our game},  each player plays the role of a fragment of the environment monitoring  $\mathcal{S}$. 
A related situation would be to consider a collisional approach to open quantum systems \cite{Ziman05-2}, but with strong interaction rather than weak. 
{Within this picture, the system first interacts with a fragment $\mathcal{E}_1$. 
This interaction is capable of establishing the right correlations, transferring information about the monitored observable-- $A_{i_1}$, say -- to the environmental subsystem. After that, the decohered system goes on to the next interaction which, by its turn, establishes new correlations to the next environmental subsystem, $\mathcal{E}_2$. See Fig. \ref{Fig: Interaction}.}

A slight modification in the game makes this interpretation {even} stronger.
In this new version, in each run the $K$ players access the system in a random order, unknown to them.
In this new version, each player would posses an average of the values of $\alpha^k$ ($\beta^k$), given by the uniform distribution of the possible results, $\bar{\alpha}_K = (1/K)\sum \alpha^k$ (similarly for $\beta$), which is now the same quantity for every player `in the environment'.  
In this case, the answer to the question `does the statistics of outcomes of measurements accept a 
non-contextual model?' would have a unique, collective, answer. 
To understand which answer this would be, 
we show  at Table \ref{Table1}, for some values of $N$ and for each protocol, the maximum number of players $K_{\text{max}}$ -- i.e., the `size of the environment' --  that still would violate the relevant inequality.
\begin{table}[ht]
\centering
\begin{tabular}{|c|c|c|c|c|c|c|} \hline
& \multicolumn{6}{c|}{$K_{max}$} \\ \hline
 & \multicolumn{3}{c|}{$K_{\text{max}}$ - Predef. Order} & \multicolumn{3}{c|}{ $K_{\text{max}}$ - Unif. Distr.} \\ \hline
$N$     & $M(\alpha , \beta) $  & $M^a(\alpha)$       & $M^b(\beta)$ & $M(\alpha , \beta)$  & $M^a(\alpha)$ & $M^b(\beta)$ \\ \hline
5 & 1 & 1 & 2 & 1 & 2 & 4 \\ 
7 & 1 & 1 & 3 & 1 & 1 & 6 \\
9 & 1 & 1 & 4 & 1 & 1 & 8 \\ 
11 & 1 & 1 & 5 & 1 & 1 & 9 \\ 
13 & 1 & 1 & 5 & 1 & 1 & 11 \\ 
15 & 1 & 1 & 6 & 1 & 1 & 12 \\ 
17 & 1 & 1 & 7 & 1 & 1 & 14 \\ 
19 & 1 & 1 & 8 & 1 & 1 & 16 \\
\hline
\end{tabular}
\caption{Number of players that can witness contextuality ($K_{max}$) for the different protocols in both versions regarding step (ii): (left) order is fixed; (right) the order is sorted randomly for the total number of players.
}
\label{Table1}
\end{table}
We can see that $K_{\text{max}}$ is always higher for protocol $3$ and for the randomized access version when compared with the other possibilities, as a consequence of the less disturbance of the measurements in this protocol. Even in this case, it is a quite small environment that can witness contextuality, for these values of $N$. 
So, our results show that \emph{contextuality is hidden for a usual-sized environment}, if $N$ is not enormously large. Even in this case, the asymptotic limit shows that there is always a big enough environment for which no-violation occur.

{One could say that our conclusion is false for enormously high values of $N$ and a fixed usual-sized environment}. Since $K_{\text{max}}$, grows with $N$ for protocol $3$, in principle, a very well designed environment could witness $N$-cycle contextuality for these values of $N$. However, the measurements involved in such high values of $N$ for $N$-cycle inequalities would require a huge precision to be performed and distinguished. We can then conclude that witnessing violation of $N$-cycle non-contextuality inequalities by having access to fragments of an (already very carefully designed) environment is very difficult, if not impossible for all practical purposes.
{\subsection{Comparison with emergence of objectivity}}

{The `Redundancy' paradigm is the stage where emergence of objectivity might be obtained. So, obtaining emergence of non-contextuality\footnote{Actually, obtaining emergence of the impossibility of attesting contextuality would be more precise, but also more cumbersome.} under this same setup calls for an analysis of the relation between our results and such different emergence of classicality. Under such a paradigm, three processes were suggested to be responsible for objectivity: QD,SSB and SQD. Historically, the process of Quantum Darwinism (QD) was the first one to be proposed \cite{ZUREKQDEnvariance,ZUREK2003}. Further development has replaced QD by State Spectrum Broadcast (SSB) structure\cite{RHorodecki_SSB} or, more recently, by Strong Quantum Darwinism (SQD)\cite{Le2019_SQD}. 
Let us concisely state the main features of these processes to then see how our results compare to the emerging classicality on them.}

In system-environment processes where QD  successfully takes place, the interaction both decoheres the system $\mathcal{S}$ in a preferred basis -- the pointer basis-- and broadcasts the information about the projection on that basis to several independent fragments of the environment\footnote{This might not always be the case in nature; for those cases a broader notion of POVM pointer states can apply \cite{BRANDAO,Beny}.}. 
This allows for independent observers to agree about this information even if they have access only to disjoint fragments of the environment, without disturbing directly the state of the system. 
{Since this information is redundantly stored in the sub-environments, having access to bigger fragments makes (almost) no difference, as long as there is some missing part.
This gives rise to the so-called \emph{classical plateau} in the mutual information plot \cite{ZUREKREVIEW2007}: (almost) irrespective of the number of environmental fragments, $\#f\mathcal{E}$, the mutual information $I(\mathcal{S}:f\mathcal{E})$ is approximately the entropy of $\mathcal{S}$.}
This whole process should be seen, according to Zurek et al.'s proposal \cite{ZUREKQDEnvariance,ZUREK2003,Ollivier_QD}, as recovering objectivity inside QT: independent observers would agree on classical information obtained about a system, without further disturbing it. 
The classical plateau being considered the signature that objectivity has indeed emerged.

{However, the classical plateau in the mutual information plot was shown insufficient to attest objectivity in the proposed sense. 
Indeed, references \cite{RHorodecki_SSB,Pleasance2017,Le2018_QDandObjectivity} have shown that this condition can be obtained by rather non-objective states. 
In these states, considerable part of the mutual information is composed of quantum correlations, instead of classical. This gave rise to another proposition: objectivity was attained when the combined state of system and environmental fragments assume the SSB structure \cite{RHorodecki_SSB}. In states with such structure, the spectrum of a pointer observable must be encoded into the fragments and no-correlations among environmental subsystems are allowed besides those established through the pointer states of $\mathcal{S}$. This structure was indeed shown \emph{sufficient} for objectivity \cite{RHorodecki_SSB}. Recently, it was shown that one can arise at \emph{necessary and sufficient} conditions for objectivity by requiring weaker conditions than SSB, through strong quantum Darwinism\cite{Le2019_SQD}. This process allows for extra correlations between environmental fragments, as long as they are not quantum correlations, i.e., as long as there is zero-discord. 

All of those processes vary in their success to signal inter-subjectivity, but they are all based in the same emergence notion of objectivity: information about a pointer basis, selected by the interaction, is spread to independent environmental fragments. 
In all of those processes, when successful, redundancy-- seen as agreement on the information gathered by independent observers-- brings classicality.}

{In the case of our game, there are already two main differences to these processes: here we consider several runs for players to collect their data and the fragments of the environment try to monitor incompatible observables. 
Interestingly, our results show that these independent players still see classicality to emerge in this program: they would, generically, be unable to attest contextuality that was initially available.
This emergence of a different classicality -- non-contextuality instead of objectivity of pointer observables -- can occur even when those processes are \emph{not} successful; in other words, an important quantum feature, quantum contextuality, can be lost even if objectivity does not prevail. 
This happens because these processes also require the selection of a pointer basis and, in each run of our game, this fails to occur since the observables monitored by each environment fragment are usually not jointly compatible. 
For instance, if one player monitors the projections $\{\ket{a_0}\bra{a_0}, \mathcal{I}-\ket{a_0}\bra{a_0}\}$, this sets a selection of classical information about $\mathcal{S}$ that might differ from the previous or the next player, if they do not choose projections commuting with $\ket{a_0}\bra{a_0}$. 
This implies that after the $i$-th interaction, correlations between $\mathcal{S}$ and the $(i-1)$-th fragments are weakened and there is no well-defined \emph{collective} classical information about $\mathcal{S}$ being broadcast to the fragments, in each run.} 
This means that there is no objectivity in this case -- the observables being monitored are not the same to begin with, so there is no reason to hope the probabilities of each `answer' to be the same. 
However, there is still emergence of non-contextuality: we can ask to the independent observers, after several runs, if the probabilities they infer about $\mathcal{S}$ allow them to violate the inequality or not. 
As discussed above, the typical answer is `no', and since this will happen for all $N$-cycle inequalities, it means that the players can agree on a non-contextual model leading to the frequencies that they recorded.
So, non-contextuality appears in such a `Redundancy program' even in a situation where objectivity does not emerge. 

As mentioned already, if we took a different approach, allowing the players to combine strategies instead of acting independently, (maximum) violation could be obtained -- e.g., if every player made the same measurement in each run (every fragment would monitor the same observable). 
This means that this emergence of non-contextuality could be frustrated  if some of the players were allowed to cooperate. 
In this system-environment interpretation, we can say that if we establish the right correlations between the environment fragments, this kind of classicality might fail to appear. 
This is similar to what happens with objectivity, which might be hindered if the right correlations are {engineered} between the environmental systems \cite{Ciampini2018}, {a rather unlikely situation}.
{Interestingly, this collective approach to the game could be exactly the case where objectivity of pointer observables would emerge \emph{in each run}, with correct information about a unique measurement spread out through the whole environment.} 
{In other words, objectivity and non-contextuality are shown independent in this game, with the former being related to consistency of outcomes in each run while the latter to the possibility of classically explain the probabilities inferred after many runs.}

\section{Conclusions}
\label{Sec: Conclusions}

In this work, we address the behaviour of quantum contextuality under strong measurements from sequential observers, in the odd $N$-cycle scenario. 
We focus on independent players for two reasons: they mimic better the situation where unknown effects act on the system and cooperative players can act as one team, bringing back the one-player usual behaviour.
We make a separate analysis for each of three protocols that are equivalent in the usual single observer setup, showing they behave differently in this new setting. 
Results point to protocol $3$ as the one that allows contextuality to survive for more players, thus being better for protecting contextuality.

Since contextuality is a striking non-classical feature, it is important to understand classical limits in which non-contextuality emerges.
It is fair to say that any classical limit which allows for contextuality is not a fully classical limit. 
By interpreting the game results in an environment-system language we obtained $N$-cycle non-contextuality as an emergent property in open quantum system approaches and compared it to {emergence of objectivity under} the `Environment as Witness' paradigm.
In other words, the impossibility of attesting contextuality emerges for most players when many \emph{independent} observers, with no collective strategies, play the game. 
The redundant conclusion among players is that it is \emph{not} possible to witness quantum contextuality.

This work also opens questions and directions for future research. 
First, it is valuable to study variations of this game that can be even more related to usual classical limits. 
For example, considering interaction between players and $\mathcal{S}$ in place of measurements, leading to an approach that is an interplay between collisional models and QD \cite{Rau63,Ciampini2018,Ziman05,Ziman05-2,Lorenzo17}. 
This scenario is also a good stage to consider weak measurements as done for non-locality \cite{Silva2015}.
It is essential to study the response of other forms of contextuality, specially \emph{state-independent} contextuality since our approach is manifestly state-dependent -- a state-recycling approach already shows state-independent contextuality might be more robust to the dynamics described here \cite{Wajs2016-StateRecycling}. 
If the kind of behaviour obtained in the present work is not present in such type of contextuality, this is interesting for quantum foundations and computation; if non-contextuality still emerges in this setting, it would be interesting to understand why, since the `state-degradation' interpretation is not possible anymore.

\begin{acknowledgments}
RDB acknowledges funding by S\~{a}o Paulo Research Foundation - FAPESP, grants no. 2016/24162-8 and no. 2019/02221-0. 
Authors thank the Brazilian agencies CNPq and CAPES for financial support.
This work is part of the Brazilian National Institute for Science and Technology in Quantum Information.
We thank Felippe Barbosa, J\'{a}nos Bergou, Raphael Drumond, Rafael Rabelo and Ana Bel\'{e}n-Sainz for fruitful discussions. 
Important clarifications were made thanks to anonymous Referees.
\end{acknowledgments}

\appendix

\onecolumngrid

\section{{Non-disturbance and `Gleason property'}}
\label{Sec: NDandGP}

In this subsection we intend to make a short exposition about some fundamental principles mentioned in the main text: non-disturbance (ND) and `Gleason property' (GP).

The ND principle can be stated as follows: a theory obey the ND principle if the marginalization of the joint distribution does not depend on the marginalized measurement. Mathematically, a simplified version considering joint measurements of only two compatible measurements, can be written
\begin{equation}
    p(o_i|i) = \sum_{o_j} p(o_i,o_j|i,j) = \sum_{o_{j'}} p(o_i,o_{j'}|i,j'),
\end{equation}
for all $A_j,A_{j'}$ compatible with $A_i$. The more general form assume similar restrictions are valid for any number of compatible measurements and any marginalization in a given context \cite{TerraBarbaraBook}. This is a way of saying that, if measurements are compatible, there should be a way of performing them such that they do not disturb the results of each other. This can be seen as a non-signaling-like principle where compatibility does not necessarily arise from spatial separation. 
If we assume ND and exclusiveness, i.e. $p(-1,-1|i,j)=0$, in a dichotomic scenario such as the $N$-cycles, one has
\begin{align}
   & p(-1|i)=p(-1,+1|i,i+1) + \cancel{p(-1,-1|i,i+1)}=p(+1,-1|i-1,i)+\cancel{p(-1,-1|i-1,i)} \,\,\forall \,i\nonumber\\
    \Rightarrow & p(-1|i)=p(-1,+1|i,i+1) =p(+1,-1|i-1,i) \,\,\forall\,i.
    \label{Eq: ExclusivenessOpEq}
\end{align}
{What Equation \eqref{Eq: ExclusivenessOpEq} says is that the events $(-1|i)$,$(-1+1|i,i+1)$ and $(+1,-1|i-1,i)$ are \emph{operationally equivalent}: any probability assigned to one of these events must be assigned to the others for the probabilistic model to respect ND and exclusiveness.} This allows us to identify the events $(-1|i), (-1,+1|i,i+1)\text{ and } (+1,-1,|i-1,i)$, and this identification allows us to take Inequalities \eqref{Ineq: ProbDif} and \eqref{Ineq: ProbEqual} to the forms
\begin{subequations}
    \begin{equation}
        \sum_i p(-1|i)\leq \frac{N-2}{2},
        \label{Ineq: pND-1}
    \end{equation}
    \begin{equation}
        \sum_i p(+1,+1|i,i+1)\geq 1.
        \label{Ineq: pND+1+1}
    \end{equation}
\end{subequations}
under the 
assumption of non-contextuality as described in the main text. {We can visualize such an identification procedure of the operationally equivalent events as identifying vertices in a hypergraph where each vertex is an event and hiperedges circles those vertices which are outcomes of a measurement. 
}

\begin{figure}[ht]
    \centering
    \includegraphics[scale=0.3]{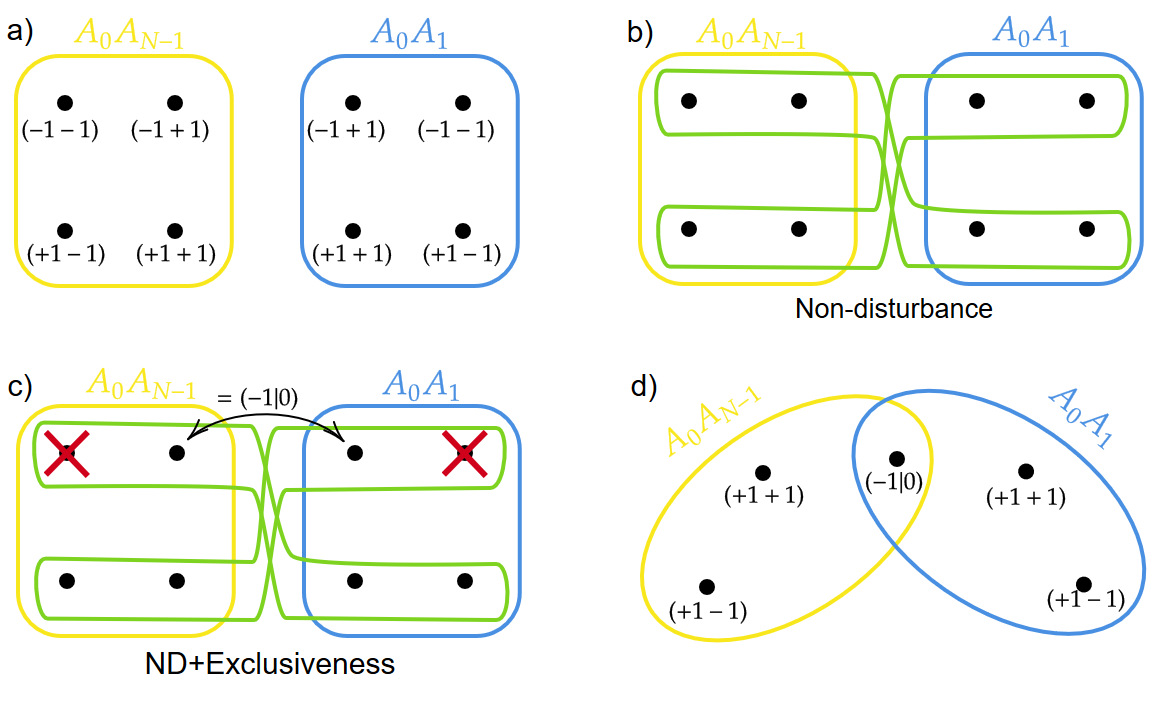}
    \caption{a) Two hyperedges defining joint measurements with its initially four outcomes. b) ND says that the probabilities assinged to the upper (lower) events on either hyperedge must sum up to the same value. This can be used together with normalization condition for each hyperedge to obtain that probabilities assinged to the events in the green hyperedges must sum up to one.c) Exclusiveness essentialy allows to take $(-1-1|i,j)$ out of consideration, implying the operation equivalence of these two vertices and the outcome of the measurement of $A_0$ alone. d) The operational equivalence allows to identify these events as one vertex and we end up with this new hypergraph for these contexts.}
    \label{Fig: HypergraphConstruction}
\end{figure}

Note that, on the above discussion, \emph{no hypothesis were made on how this probabilities arise}: how states are represented or how measurements act on them, etc.. 

However, one can also consider some assumptions on the mechanism that lead to such distributions. Since we analyze and compare different measurement protocols, it is also important in this work to consider such additional structure for general theories, the proper language for contextuality scenarios, to only then specialize it to QT. A broad and interesting framework is the one given by generalized probabilistic theories (GPTs) \cite{Barret2007,Janotta2014}. In this framework, one usually considers that propositions about physical systems can be represented by linear functionals called effects, acting on the set of states. Most importantly, effects are in a one-to-one relation to an equivalence class of propositions/events, that is, if two outcomes are attributed the same probability when acting on an initial state (for all initial states), these outcomes are represented by the same effect \cite{Janotta2014}.
The effects can then be combined to form complete measurements $\{e_i\}$, summing up to a special functional $u$ that gives probability one to all normalized states, $\sum e_i=u$.

All this structure which considers (equivalence classes of) propositions to obtain probabilities, irrespective of the rest of the measurement, implies that the effects cannot depend on the others with which they are combined to give a measurement. This is what is called `Gleason Property' \cite{Cabello2010}.
This applies to our case directly: since we can identify $(-1|i),(-1,+1|i,i+1)$ and $(+1,-1|i-1,i)$ these outcomes \emph{are represented by the same effect, namely $a_i$} by any probabilistic model obeying GP -- in particular, the usual approach on GPTs. This means that a joint measurement $A_iA_{i+1}$ for measurements $A_i=\{a_i,\neg a_i\}$ must be made of $a_i$ itself, $a_{i+1}$ and a third effect, which we denote by $b_i$. In other words, a joint measurement $A_0A_1$ that would initially be made of effects $\{a_0\wedge a_1,a_0\wedge\neg a_1,\neg a_0\wedge a_1, \neg a_0\wedge \neg a_1\}$ is adapted, given the operational equivalence brought by exclusiveness and ND, by identifying $a_0\leftrightarrow a_0\wedge\neg a_1$,$a_1\leftrightarrow\neg a-0\wedge a_1$ and $a_0\wedge a_1\leftrightarrow 0$. Finally, we just rename $b_0\equiv \neg a_0\wedge \neg a_1$.
This identification can be in every context to get to inequalities \eqref{Ineq: ProbAeB} and to adapt straightforwardly the hypergraph in Figure \ref{Fig: HypergraphConstruction} where vertices are effects and hyperedges show those that form a measurement. When considering all the contexts, one gets a hypergraph like the one in Fig. \ref{Fig: HyperGraph} for $N=5$.

\section{Protocol number 1}
\label{Sec: Protocol1}

In protocol number $1$, the measurements are given by the set of complete projective measurements $\{\ket{a_i}\bra{a_i},\ket{b_i}\bra{b_i},\ket{a_{i+1}}\bra{a_{i+1}}\}_i$, where $i\in\{0,...,N-1\}$ denotes the $N$ possible choices and the sets of outcomes are $\{a_i,b_i,a_{i+1}\}_i$ \footnote{It is important to stress that this notation is related to the GP assumption. It is used in denoting equivalence between the \emph{events} $a_{i_k+1}|i_k = a_{i_k+1}|i_{k+1}$, and is consistent with the use of the same projector in both measurements 
within Quantum Theory.}. We can define the  vector $\vec{P}_{i_k}$ for the $k$-th observer ($i_k$ labeling the choice of measurement for this player) as:
\begin{align}
    \vec{P}_{i_k} =\left(\begin{array}{c}
    p(a_{i_k}|i_k)\\
    p(b_{i_k}|i_k)\\
    p(a_{i_k+1}|i_k)\end{array}\right),
\end{align}
where the entries define probabilities for the different outcomes of the measurement. Using the definitions for $\alpha^k$ and $\beta^k$, given by \eqref{Ineq: Prob}, we can write the sums involved in terms of this vector, obtaining the following:
\begin{subequations}
\begin{align}
 &\beta^k=\sum_{i_k}\left(\begin{array}{ccc}0 &1& 0 \end{array}\right) \vec{P}_{i_k} = \left(\begin{array}{ccc}0 &1& 0 \end{array}\right) \sum_{i_k}\vec{P}_{i_k} 
 \label{Eq: BetaProtocol1}\\
 &\alpha^k = \sum_{i_k}\left(\begin{array}{ccc}1 &0& 0 \end{array}\right) \vec{P}_{i_k} = \sum_{i_k}\frac{1}{2}\left(\begin{array}{ccc}1 &0& 1 \end{array}\right) \vec{P}_{i_k}=
 \frac{1}{2}\left(\begin{array}{ccc}1 &0& 1 \end{array}\right) \sum_{i_k}\vec{P}_{i_k}.
 \label{Eq: GammaProtocol1}
 \end{align}
 \label{Eq: Protocol1}
 \end{subequations}
Some comments on \eqref{Eq: GammaProtocol1} are important: the original expression for $\alpha$, Eq. (3a) of the main work, suggests the inner product of each $\vec{P}_{i_k}$ with $(1,0,0)^T$ which picks out the value $p(a_{i_k}|i_k) = \langle \ket{a_{i_k}}\bra{a_{i_k}} \rangle$. However, the last expression is also correct under the non-disturbance $+$ exclusiveness conditions \cite{TerraBarbaraBook}, which is valid for quantum theory and can be stated in  this case as:
  \begin{equation}
     p\left(a_{i_k+1}|i_k\right) = p\left(a_{i_k+1}|i_k+1\right), \,\, \forall i_k,i_{k+1}.
\end{equation}
This means, in vector form:
 \begin{equation}
     \left(\begin{array}{ccc}0& 0& 1\end{array}\right)\vec{P}_{i_k} = \left(\begin{array}{ccc}1& 0& 0\end{array}\right)\vec{P}_{i_k+1},
 \end{equation}
which in turn implies that any convex combination of these entries gives the same; in particular, the right hand side of expression \eqref{Eq: GammaProtocol1}. We chose this particular convex combination because of its simplicity and symmetry. These forms for $\alpha^k$ and $\beta^k$ can be summarized by the expression below:
\begin{equation}
  I^k = \sum_{i_k}\langle v_{I},\vec{P}_{i_k}\rangle,
  \label{Eq: IPk}
\end{equation}
where $I$ represent either $\alpha$ or $\beta$ and $v_I$ is respectively $(1/2)(1 \,\, 0 \,\, 1)$  or $(1 \,\, 0 \,\, 0)$. We can understand Expression \eqref{Eq: IPk} as follows: for every observer $k$, the collection $(P_{i_k})_{i_k}$ that maximizes $\alpha^k$ is the one that maximizes the extreme entries of $\sum P_{i_k}$ (and with the similar understanding for minimizing $\beta^k$).

Given Expressions \eqref{Eq: Protocol1} for $\beta^k$ and $\alpha^k$ in terms of the vectors $\vec{P}_{i_k}$ we can state the problem as a Markovian process, as follows.
Denoting as $\rho^k$ the state that is available to the $k$-th player after the measurements of the precedent ones and denoting either $\{a_{i_k},b_{i_k},a_{i_k+1}\}$ as $o_{i_k}$ we can write
\begin{align}
    p(o_{i_k}|i_k) &= \bra{o_{i_k}}\rho^k\ket{o_{i_k}}= \Tr\left[\ket{o_{i_k}}\bra{o_{i_k}}\rho^k\ket{o_{i_k}}\bra{o_{i_k}}\right] \nonumber \\ &=\frac{1}{N}\sum_{ i_{k-1},o_{i_k-1}}\Tr\left[\ket{o_{i_k}}\braket{o_{i_k}}{o_{i_{k-1}}}\bra{o_{i_{k-1}}}\rho^{k-1}\ket{o_{i_{k-1}}}\braket{o_{i_{k-1}}}{o_{i_k}}\bra{o_{i_k}}\right] \nonumber\\ &=\frac{1}{N}\sum_{i_{k-1},o_{i_k-1}} |\braket{o_{i_k}}{o_{i_{k-1}}}|^2p(o_{i_{k-1}}|i_{k-1}).
\end{align}
We used the fact that the players choose the measurement independently and from an uniform distribution under the $N$ possible choices for defining $\rho^{k-1}$. We can rewrite the relation above in matrix form as 
\begin{align}
    \vec{P}_{i_k} =\frac{1}{N}\sum_{i_{k-1}=0}^{N-1}\left( \begin{array}{ccc}
    |\braket{a_{i_{k-1}}}{a_{i_k}}|^2 & |\braket{b_{i_{k-1}}}{a_{i_k}}|^2 & |\braket{a_{i_{k-1}+1}}{a_{i_k}}|^2\\
    |\braket{a_{i_{k-1}}}{b_{i_k}}|^2 & |\braket{b_{i_{k-1}}}{b_{i_k}}|^2 & |\braket{a_{i_{k-1}+1}}{b_{i_k}}|^2\\
    |\braket{a_{i_{k-1}}}{a_{i_k+1}}|^2 & |\braket{b_{i_{k-1}}}{a_{i_k+1}}|^2 & |\braket{a_{i_{k-1}+1}}{a_{i_k+1}}|^2\\
    \end{array}\right)\vec{P}_{i_{k-1}} =\frac{1}{N}\sum_{i_{k-1}=0}^{N-1} \left[M_{i_k,i_{k-1}}\right]\vec{P}_{i_{k-1}},
    \label{MyxPx}
\end{align}
which defines the matrix $M_{i_k,i_{k-1}}$: each column is labeled by a fixed vector representing an outcome $o_{i_{k-1}}|i_{k-1}$ of the $k-1$-th observer, while the rows are labeled by vectors representing the outcomes $o_{i_k}|i_k$ of the $k$-th observer. It is important to note that, once $N$ is fixed, this matrix depends only on the difference $i_k - i_{k-1}$ (modulo $N$) and is bistochastic --  rows and columns sum to one as they are the norm of normalized states-- denoting the Markovianity of this process. Indeed, if the choice $i_{k-1}$ and values of $\vec{P}_{i_{k-1}}$ are known, then the whole vector $\vec{P}_{i_k}$ can be obtained.

Now we are going to investigate the consequences of such Markovianity, leading to the results for this protocol. Using \eqref{MyxPx} in Eq. \eqref{Eq: IPk} we can obtain the expression for $I^k$ in terms of the matrices $\{M_{i_ki_{k-1}}\}$:
\begin{align}
I^k = \sum_{i_k}\left\langle v_I\,,\, \vec{P}_{i_k} \right\rangle= \sum_{i_k}\left\langle v_I,\frac{1}{N}\sum_{i_{k-1}} M_{i_k,i_{k-1}}\vec{P}_{i_{k-1}}\right\rangle 
 = \left\langle v_I\,,\,\sum_{i_{k-1}}\left(\frac{1}{N}\sum_{i_k} M_{i_k,i_{k-1}}\right)\vec{P}_{i_{k-1}}\right \rangle.
\label{Eq: OperatorM}
\end{align}
The term in parenthesis suggests the definition of the matrix
\begin{equation}
    \mathbb{M}_N = \frac{1}{N}\sum_{i_k} M_{i_k,i_{k-1}},
    \label{Eq: DefMN}
\end{equation}
which does not depend on $i_{k-1}$, since each $M_{i_k,i_{k-1}}$ depend only on the distance $i_k - i_{k-1}$ and, by summing $M_{i_k,i_{k-1}}$ over $i_k$, all differences $i_k-i_{k-1}$ are considered. We can see from Eq. \eqref{Eq: DefMN} that $\mathbb{M}_N$ is also bistochastic. The symmetry of the $N$-cycles and the bistochastic feature reflect into $\mathbb{M}_N$ not only by making it independent of $i_{k-1}$ but also imply that  $(\mathbb{M})_{11}=(\mathbb{M})_{13}=(\mathbb{M})_{31}=(\mathbb{M})_{33}\equiv t_N$ and $(\mathbb{M})_{12}=(\mathbb{M})_{21}=(\mathbb{M})_{23}=(\mathbb{M})_{32} = 1-2t_N$ (see Eq. \eqref{MyxPx}). So, we can define $\mathbb{M}_N$ only in terms of $t_N$ as:
\begin{align}
   \mathbb{M}_N =\left( \begin{array}{ccc}
    t_N & 1-2t_N & t_N \\ 
    1-2t_N & 4t_N - 1 & 1-2t_N\\
    t_N & 1-2t_N & t_N
    \end{array}\right).
    \label{Eq: MNSymmetry}
\end{align}
With $M_N$ in this form we can get bounds on $t_N$, since all elements being positive implies $1\leq4t_N\leq 2$ for all odd $N$.

As we know the important features of the matrix $\mathbb{M}_N$ we can turn back to its effects on the inequalities for the $k$-th player. From Eq. \eqref{Eq: OperatorM} and the symmetries of $\mathbb{M}_N$, we can write:
\begin{equation}
I^k = \left\langle v_I\,,\,\sum_{i_{k-1}}\mathbb{M}_N\vec{P}_{i_{k-1}}\right \rangle=  \left\langle v_I\,,\,\mathbb{M}_N\sum_{i_{k-1}}\vec{P}_{i_{k-1}}\right \rangle . 
\label{Eq: IneqMN}
\end{equation}
Equation \eqref{Eq: IneqMN} highlights the Markovianity of the game under protocol number $1$. We can see that the relevant vector for the $k$-th player to calculate $I_k$ --i.e., $\sum\vec{P}_{i_k}$ --  can be obtained only from $\mathbb{M}_N$ and the relevant vector of the previous player, $\sum\vec{P}_{i_{k-1}}$.
Using the same reasoning as in Eq. \eqref{Eq: OperatorM} for $\vec{P}_{i_{k-1}}$ in terms of $\vec{P}_{i_{k-2}}$ and so on, we arrive at
\begin{equation}
   I^k= \left\langle v_I,\mathbb{M}_N\sum_{i_{k-1}}\vec{P}_{i_{k-1}}\right\rangle=\left\langle v_I,(\mathbb{M}_N)^{k-1}\sum_{i_{1}}\vec{P}_{i_{1}}\right\rangle.
   \label{Eq: OperatorMN}
\end{equation}
Since the matrix $\mathbb{M}_N$ is fixed for each $N$, Eq. \eqref{Eq: OperatorMN} already gives all that is necessary to calculate $\alpha^k$ and $\beta^k$ for any $k$ and initial state in terms of $(P_{i_1})_{i_1}$.

With Eqs. \eqref{Eq: MNSymmetry},\eqref{Eq: IneqMN} and \eqref{Eq: OperatorMN} it is possible to prove that the initial state that gives maximal violation for the first player also allows for the best attempt for the $k$-th player to win. The proof goes as follows. Writing $\sum_{i_{k-1}}\vec{P}_{i_{k-1}} = (c\,\,d\,\,e)^t$ for the $(k-1)$-th player, $\alpha^{k-1}$ is maximized ($\beta^{k-1}$ is minimized) by the vector with maximum $(c+e)$, as pointed above \footnote{For Quantum Theory (and for any theory respecting the non-disturbance principle) $c=e$, since $p(a_j|i_j) = p(a_j|i_{j-1}) \,\,\forall j$ imply that the sums of the extreme entries are equivalent. However, we do not need to consider it here.}. Now, if this vector is multiplied by $\mathbb{M}_N$, from Eq. \eqref{Eq: MNSymmetry} we have
\begin{equation}
   \sum_{i_k}\vec{P}_{i_k}= \mathbb{M}_N\sum_{i_{k-1}}\vec{P}_{i_{k-1}} = (c+e)\left(\begin{array}{c} t_N \\1-2t_N \\ t_N\end{array}\right) + d\left(\begin{array}{c}1-2t_N\\4t_N - 1\\1-2t_N \end{array}\right).
\end{equation}
Now, $\alpha^k$ can be written as 
\begin{equation}
   \alpha^k= (c+e)t_N + d(1-2t_N) = (c+e)(3t_N-1)+N(1-2t_N),
\end{equation}
where we used that $c+d+e=N$. We know from the positivity of every element of $\mathbb{M}
_N$ that $t_N<1/2$, ensuring positivity of the last term, $(1-2t_N)$. Now, we see that the maximum value of $\alpha^k$ will be given by the maximum value of $(c+e)$ only if $t_N>1/3$. For those cases, we see that the vector $\sum P_{i_{k-1}}$ that maximizes $\alpha^{k-1}$ (the one with maximum $(c+e)$) leads to the vector $\sum P_{i_k}$ that maximizes $\alpha^k$, i.e., leads to the vector $\sum_{i_k}\vec{P}_{i_k}$ with maximized extreme entries. Then, continuing the argument for the previous players, the vector $\sum P_{i_1}$ that leads to the best value of $I^k$ is the one that leads to the best value for $I^1$. Since we know that the quantum state $\ket{\Psi_{\text{handle}}}$ leads to maximum violation for the first player, we know by the previous argument that this initial state also leads to the best hope for the $k$-th player to win, for all $k$. It is left to prove now that $t_N>1/3$ is not a restriction in the scenarios we are considering. In fact, it follows from (i) $t_N = (1/ N)\sum|\braket{a_0}{a_i}|^2$ increases with $N$ and (ii) $t_{N=3}=1/3$, implying that $t_N>1/3 \,\, \forall \,\, N\geq 5$ \footnote{For $N=3$, two out of the three terms in the sum are zero, since each state is orthogonal to the other two, and the remaining term is $1$; this implies $t_3=\mathbb{M}_{11}=1/3$.}.

The calculation given by Eq. \eqref{Eq: OperatorMN} can be used for every $k$ and leads naturally to the evaluation of the limit $k \rightarrow \infty$. The asymptotic limit can be calculated exaclty by noting that the matrix $\mathbb{M}_N$ is regular (which implies ireducibility). A stochastic matrix $M$ is said to be regular if there is a  natural $r$ such that all entries of the $r$-th power of $M$ are positive:
\begin{equation}
    (M^r)_{ij} > 0, \forall i,j.
\end{equation}

For the matrices $\mathbb{M}_N$ it is possible to see that $r=1$. This is so because there are null entries in $M_{i_k,i_{k-1}}$ if, and only if, $|i_k - i_{k-1}|\leq 1$. In all other cases, the definition of the vectors $\ket{o_{i_k}}$ imply strictly positive entries for $M_{{i_k},i_{k-1}}$. As $\mathbb{M}_N$ is the sum of all elements of $\{M_{{i_k},{i_{k-1}}}\}$, $\mathbb{M}_N$ has all entries strictly higher than zero.  Now, the Perron-Fr\"{o}benius theorem guarantees the existence of a unique stationary distribution eigenvector (i.e., eigenvector with eigenvalue $1$) for any regular stochastic matrix, for which the system will tend by successive applications of this matrix \cite{DurrettStochasticBook}. This in turn implies that $\mathbb{M}_N$ has such an unique stationary distribution; more than that, as $\mathbb{M}_N$ is not only stochastic but bistochastic, it is a fact that the uniform distribution $\vec{P}^* = (1/3)(1,1,1)^T$ is an eigenvector with eigenvalue $1$. As the Perron-Fr\"{o}benius theorem guarantees the stationary vector is unique, this is the only stationary distribution. In other words:
\begin{equation}
    \lim_{n\rightarrow\infty} \left(\mathbb{M}_N\right)^n \vec{P} =\frac{1}{3} \left(\begin{array}{c}
    1\\
    1\\
    1\end{array}\right), \,\, \forall \vec{P} \text{  s. t.}\,\, \sum_i P(i)=1.
\end{equation}
This implies the asymptotic limit presented in the text.

\section{Protocols number 2 and 3}
\label{Sec: Protocols2and3}

For protocols $2$ and $3$, the measurements are $\{\ket{a_{i_k}}\bra{a_{i_k}},\mathcal{I}-\ket{a_{i_k}}\bra{a_{i_k}}\}_{i_k}$ with outcomes $o_{i_k}|i_k\in \{a_{i_k},\neg a_{i_k}\}$ and  $\{\ket{b_{i_k}}\bra{b_{i_k}},\mathcal{I}-\ket{b_{i_k}}\bra{b_{i_k}}\}_{i_k}$ with outcomes $o_{i_k}|i_k\in \{b_{i_k},\neg b_{i_k}\}$, respectively. Here we prove \eqref{Eq: IneqProtocols2e3}, relating $\beta^k$ to $\beta^{k-1}$ and $\alpha^k$ to  $\alpha^{k-1}$:
\begin{subequations}
\begin{align}
    \beta^k &= B_N\beta^{k-1}+b_N,\\
    \alpha^k &= C_N\alpha^{k-1} +c_N.
    \end{align}
\end{subequations}
We will present the calculations for $\beta^k$ and the results for $\alpha^k$ follows an analogous path. 

Expression for $\beta^k$ can be written in the form
\begin{equation}
    \beta^k = \sum_{i_k}\Tr\left\{\ket{b_{i_k}}{\bra{b_{i_k}}}\rho^k\ket{b_{i_k}}{\bra{b_{i_k}}}\right\} = \Tr\left\{\left(\sum_{i_k}\ket{b_{i_k}}{\bra{b_{i_k}}}\right)\rho^k\right\} = \Tr\left\{\mathcal{B}_N \rho^k\right\}
\end{equation}
with $\mathcal{B}_N = \sum_{i_k}\ket{b_{i_k}}\bra{b_{i_k}}$, and $\rho^k$ the state available for the $k$-th player. Since this state is given by the initial state $\rho^1$ measured by the $k-1$ previous players without registering of which measurement and which result occurred, $\rho^k$ is given by:
\begin{equation}
    \rho^k = \left(\frac{1}{N}\right)^{k-1}\sum_{\vec{i},\vec{o}}\Pi_{i_{k-1}}^{o_{i_{k-1}}}...\Pi_{i_1}^{o_{i_1}}\rho\Pi_{i_1}^{o_{i_1}}...\Pi_{i_{k-1}}^{o_{i_{k-1}}},
\end{equation}
where we denoted the projective elements of the POVM $\{\ket{b_{i_k}}\bra{b_{i_K}},\mathcal{I}-\ket{b_{i_k}}\bra{b_{i_K}}\}$ as $\{\Pi_{i_k}^{o_{i_k}}\}_{o_{i_k}}$ with $o_{i_k}\in\{b_{i_k},\neg b_{i_k}\}$, $\vec{o}$ stands for $o_1|i_1,...,o_{i_{k-1}}|i_{k-1}$ while $\vec{i} = i_1,...,i_{k-1}$ below the sum. As one can see, the operator $\mathcal{B}_N$ defines the expression $\beta^k$ through its expected value on a given state $\rho^k$. Now, some symmetry aspects of such operator (and the analogous for the operator for $\alpha$) will be discussed. 

\subsection{Symmetries of the odd N-cycle}
\label{Sec: AssympState}
The symmetry of the vectors defining the quantum realization leads to some important relations. 
First, the action of a rotation of $2\pi/N$ by an axis defined by the $\ket{\psi_{handle}}$ or the reflection by a plane that contains both the handle and one of the vectors let the set of operators $\{\ket{b_{i_k}}\bra{b_{i_k}}\}_{i_k}$ unchanged, since it only reorganize its elements \footnote{The $N$-cycle scenario itself has a different symmetry group, since a rotation by $\pi$ along an axis on the plane passing trhough any of the vertices of the $C_N$ compatibility graph is also an element of the group. However, the special use of the third dimension for the quantum realization prohibits this element of the symmetry.}. This means that the operator $\mathcal{B}_N$ is the same after action of any element of such symmetry group (denoted $C_{Nv}$ point group in Ref. \cite{InuiTanabeOnoderaBook}). In other words, for any unitary $U_N$ representing any of the elements of the group, the following relation is valid:
\begin{equation}
    U_N \mathcal{B}_N U_N^{-1}= \sum_i U_N\ket{b_i}\bra{b_i}U_N^{-1} = \mathcal{B}_N \Rightarrow \left[\mathcal{B}_N,U_N\right]=0\,\,\forall\,\, U_N \in C_N.
\end{equation}
where we used subscript $i$ instead of $i_k$ for sake of an enlightened notation and since this is valid for all $k$.
If the representations given by $\{U_N\}$ were irreducible, Schur's lemma would imply that the operator $\mathcal{B}_N$ be a multiple of identity. However, this is not the case, as none of the groups $C_{Nv}$, with odd $N$ has irreducible representations of dimension $3$ \cite{InuiTanabeOnoderaBook}. However, these do have irreducible representation of dimension $2$.
This implies, by Schur's Lemma, that this matrix takes the form:
\begin{equation}
    \mathcal{B}_N = \sum_i\ket{b_i}\bra{b_i} = \lambda_0^N\mathcal{I}_{2\times2}\oplus\lambda_1^N\mathcal{I}_{1 \times 1}.
    \label{Eq: DiagformOn}
\end{equation}
Of course the same happens with operator $\mathcal{A}_N = \sum_i \ket{a_i}\bra{a_i}$, which acquires the form
\begin{equation}
    \mathcal{A}_N = \sum_i\ket{a_i}\bra{a_i} = \mu_0^N\mathcal{I}_{2\times2}\oplus\mu_1^N \mathcal{I}_{1 \times 1}.
    \label{Eq: DiagFormPn}
\end{equation}

\subsection{Proof for protocols 2 and 3}

The symmetry relation $\eqref{Eq: DiagformOn}$, implies the operator identity
\begin{equation}
    \sum_{i}\Pi^{b_i}_{i}\mathcal{B}_N\Pi^{b_i}_{i} = z_N\mathcal{B}_N;
    \label{Eq: PibOn}
\end{equation}
which says that when $\Pi^{b_i}_{i}$ acts on $\mathcal{B}_N$ we get a multiple of the same $\mathcal{B}_N$, where the constant of proportionality is given by $z_N = (1/N)[2(\lambda_0^N)^2 + (\lambda_1^N)^2]$ and is dependent only on the scenario, i.e., on the odd $N$. It is also possible to analyze the action of $\Pi^{\neg b_i}_i = \mathcal{I} - \Pi_i^{b_i}$ on the same operator, and one gets: 
\begin{equation}
    \sum_{i}\Pi^{\neg b_i}_{i}\mathcal{B}_N\Pi^{\neg b_i}_{i} = u_N\mathcal{B}_N + 2\lambda^N_0\lambda^N_1\mathcal{I}_{3\times3},
    \label{Eq: PiNegbOn}
\end{equation}
where $u_N = N + z_N - 2(\lambda_0^N + \lambda_1^N)$, $\lambda_0^N$ and $\lambda_1^N$ are the eigenvalues of $\mathcal{B}_N$ and so $u_N$ is also only dependent on $N$. Now, it is good to rewrite the expressions for $\beta^k$  making explicit the dependence on $\rho^{k-1}$: 
\begin{equation}
\beta^k = \sum_{i_k}p(o_{i_k}=b|i_k)=\Tr\left\{\mathcal{B}_N\rho^k\right\} = \left(\frac{1}{N}\right)\sum_{i_{k-1},o_{i_{k-1}}}\Tr\left\{\mathcal{B}_N\left(\Pi_{i_{k-1}}^{o_{i_{k-1}}}\rho^{k-1}\Pi_{i_{k-1}}^{o_{i_{k-1}}}\right)\right\}.
\label{Eq: Betakrhokm1}
\end{equation}
Using in Eq. \eqref{Eq: Betakrhokm1} the relations \eqref{Eq: PibOn} and \eqref{Eq: PiNegbOn} together with the cyclic property of the trace operation, we get:
\begin{equation}
\beta^k = \frac{z_N+u_N}{N}\beta^{k-1} + 2\frac{\lambda_o^N\lambda_1^N}{N} = B_N\beta^{k-1} + b_N,
\label{Eq: BetakPrevious}
\end{equation}
which is the relation we aimed to prove. For the asymptotic limit, we use the explicit form for $u_N$ and $z_N$, to get
\begin{equation}
    B_N = 1-\frac{3b_N}{N}.
\end{equation}
Now, we just have to calculate the limit
\begin{equation}
    \lim_{k\rightarrow \infty}\beta^k =  \lim_{k\rightarrow \infty} \left(\sum_{n=0}^k \left[B_N\right]^n\right)b_N = \frac{1}{1-B_N}b_N = \frac{N}{3},
\end{equation}
where the fact that $|B_N|<1$ for all odd $N$ was used. This finishes the proof for protocol $3$, and for protocol $2$ the proof is completely analogous. Now it is only left to prove that the average asymptotic state is the maximally mixed, for all the protocols.

\section{Average state in the asymptotic limit}

To see that the average state is indeed the maximally mixed, we first note that for this state $p(o=b_i|i) = (1/3)$ \emph{for all choices of $i$}. This implies that $\bar{\rho}_{\infty}$ can be written as
\begin{equation}
    \bar{\rho}_{\infty} = \frac{1}{3}\ket{b_i}\bra{b_i} + \frac{2}{3}\Pi_{i}^{\neg b_i}R_i\Pi_{i}^{\neg b_i}, \,\,\, \forall i \in\{0,...N-1\}
    \label{Eq: averageSBi}
\end{equation}
with $R_i$ being a positive semidefinite matrix with unit trace. Now,
\begin{equation}
    \bar{\rho}_{\infty} = \frac{N\bar{\rho}_{\infty}}{N} = \frac{1}{N}\sum_i\left(\frac{1}{3}\ket{b_i}\bra{b_i} + \frac{2}{3}\Pi_{i}^{\neg b_i}R_i\Pi_{i}^{\neg b_i}\right) =\frac{1}{N}\left(\frac{1}{3}\mathcal{B}_N + \frac{2}{3}\sum_i\Pi_{i}^{\neg b_i}R_i\Pi_{i}^{\neg b_i}\right).
\end{equation}
Above we used the fact that Equation~\eqref{Eq: averageSBi} is valid for all $i$. The matrix $\sum_i \Pi_i^{\neg b_i} R_i \Pi_i^{\neg b_i}$ also commutes with every operator representing the group $C_{Nv}$. Then, by Schur's lemma this means
\begin{equation}
    \sum_i\Pi_{i}^{\neg b_i}R_i\Pi_{i}^{\neg b_i} = \left(r_1\mathcal{I}_{2\times2}\right)\oplus r_0 \mathcal{I}_{1 \times 1} .
\end{equation}
We also showed, in Eq. \eqref{Eq: DiagformOn} that $\mathcal{B}_N$ has an analogous form. This means that the average state also has this form and we can write
\begin{equation}
    \bar{\rho}_{\infty} =\left(r'_1 \mathcal{I}_{2\times 2}\right)\oplus  r'_0\mathcal{I}_{1 \times 1},
    \label{Eq: averageSum}
\end{equation}
with $r'_j$ related to the eigenvalues of $\mathcal{B}_N$ and to $r_j$. However, it is not necessary to enter into these details. By \eqref{Eq: averageSBi}, \eqref{Eq: averageSum} and the orthogonality of $\Pi_i^{\neg b_i}R_i\Pi_i^{\neg b_i}$ with $\ket{b_i}\bra{b_i}$, the average state is a diagonal matrix,
with one eigenvalue equal to $r'_0= 1/3$ and two
other eigenvalues equal to $r'_1$. By the condition of unit trace for $\bar{\rho}_\infty$ this means that $r'_1 = 1/3$ as well.This implies $\bar{\rho}_{\infty}=\mathcal{I}/3$.

\bibliographystyle{apsrev4-1}
\bibliography{Bibliography}
\end{document}